\documentclass[aps,prl,twocolumn,superscriptaddress,longbibliography,reprint]{revtex4-2}

\usepackage{amsmath}
\usepackage{amssymb}
\usepackage{float}
\usepackage{graphicx}
\usepackage{subfigure}
\usepackage{color}
\usepackage{bbold}
\usepackage{hyperref}

\newcommand{\ie}{{\textit{i.e. }}}

\newcommand{\vs}{\textit{versus }}

\newcommand{\trace}[1]{{\mathrm{Tr}\left[#1\right]}}
\newcommand{\id}{{\mathbb{1}}}
\newcommand{\bra}[1]{\langle #1 |}
\newcommand{\ket}[1]{|#1\rangle}

\newcommand{\ketbra}[2]{\ket{#1}\hspace{-.05cm}\bra{#2}}
\newcommand{\expctval}[3]{\langle #1 | #2 | #3 \rangle}
\newcommand{\uu}{\uparrow \uparrow}
\newcommand{\dd}{\downarrow \downarrow}

\newcommand{\sx}{\hat{\sigma}^x}
\newcommand{\sy}{\hat{\sigma}^y}
\newcommand{\sz}{\hat{\sigma}^z}

\newcommand{\Hop}{\hat{H}}
\newcommand{\dimH}{dim\{\mathcal{H}\}}
\newcommand{\nstates}{K}
\newcommand{\bd}{M}
\newcommand{\GHZ}{\mathrm{GHZ}}
\newcommand{\osmall}{\mathcal{O}}
\newcommand{\eoff}{EoF}
\newcommand{\eof}{EoF }
\newcommand{\Svn}{\mathcal{S}}
\definecolor{smoothred}{rgb}{0.77,0.14,0.19}
\definecolor{mygreen}{rgb}{0,0.5,0}
\definecolor{myblue}{rgb}{0,0,0.75}
\definecolor{mymagenta}{rgb}{0.75,0.20,0.75}
\definecolor{cobalt}{rgb}{0,0.28,0.67}

\graphicspath{{Figures/}{SM_Figures/}}

\begin{document}
%
%
\title{Entanglement of formation of mixed many-body quantum states via Tree Tensor Operators}
\author{L. Arceci}
\affiliation{Dipartimento  di  Fisica  e  Astronomia  “G.  Galilei”,  Universit\`a  di  Padova,  I-35131  Padova,  Italy}
\affiliation{INFN,  Sezione  di  Padova,  I-35131  Padova,  Italy}
\author{P. Silvi}
\affiliation{Center for Quantum Physics, Faculty of Mathematics, Computer Science and Physics, University of Innsbruck, A-6020, Innsbruck, Austria}
\affiliation{Institute for Quantum Optics and Quantum Information of the Austrian Academy of Sciences, A-6020 Innsbruck, Austria}
\author{S. Montangero}
\affiliation{Dipartimento  di  Fisica  e  Astronomia  “G.  Galilei”,  Universit\`a  di  Padova,  I-35131  Padova,  Italy}
\affiliation{INFN,  Sezione  di  Padova,  I-35131  Padova,  Italy}
\begin{abstract}
We present a numerical strategy to efficiently estimate bipartite entanglement measures, and in particular the Entanglement of Formation, for many-body quantum systems on a lattice. Our approach exploits 
the \textit{Tree Tensor Operator} tensor network ansatz, a positive loopless representation for density matrices which, as we demonstrate, efficiently encodes information on bipartite entanglement, enabling the up-scaling of entanglement estimation. 
Employing this technique, we observe a finite-size scaling law for the entanglement of formation in 1D critical lattice models at finite temperature for up to 128 spins, extending to mixed states the scaling law for the entanglement entropy.
\end{abstract}
\maketitle

Quantum entanglement, correlations uniquely present in quantum systems \cite{Horodecki_RMP09},
lies at the heart of the second quantum revolution. It is a fundamental resource in the development of present and future quantum technologies \cite{MacFarlane_PTRSL03},
and it drives the collective physics of many-body quantum systems at low temperatures \cite{Qinfo_meets_Qmatter,Srednicki93arealaw}.
The ability to characterize and quantify entanglement in a quantum state is thus crucial.
However, even the simplest entanglement characterization, {\it bipartite} entanglement $-$ quantifying the mutual quantum correlations between two subsystems $-$ is well-understood only when the state of the joint subsystems is 
a {\it pure} quantum state. This is mostly due to the fact that the estimation strategies for entanglement of mixed states call for 
minimizations in spaces that scales exponentially with the number of constituents of the system, and thus 
are effectively limited to small-sized systems \cite{Loss_Elsevier11,Mintert_PR05}.
In this letter, we show how tensor network (TN) techniques can tackle this challenge, and efficiently estimate the Entanglement of Formation (\eoff)~\cite{Plenio_ReviewEntanglement} $-$ the convex-roof extension of the Von Neumann entropy $-$ of many-body quantum states. As first application of this approach, we show that for critical one-dimensional systems the EoF obeys a (logarithmic) finite-size conformal scaling-law, for temperatures commensurate with the energy gap.

For pure states, the connection between bipartite entanglement and the effective entropy of either subsystem has been largely established, and is typically expressed in terms of Von Neumann ($\Svn$) or R{\'e}nyi entropies \cite{Bruss_JMP02,Plenio_ReviewEntanglement,Renyi61,RenyiMeasurable}.
While challenging to measure in an experiment \cite{RenyiExperimentMeasure}, these estimators are often accessible in numerical simulations of many-body quantum systems, and especially in loopless tensor network ansatz states, where the calculation complexity scales polinomially with the system size
\cite{MPSReps,Uli_ageofMPS,VidalTTN2D,GersterTTN,Silvi_Anthology19}.
\begin{figure}
	\centering
	\includegraphics[width=\columnwidth]{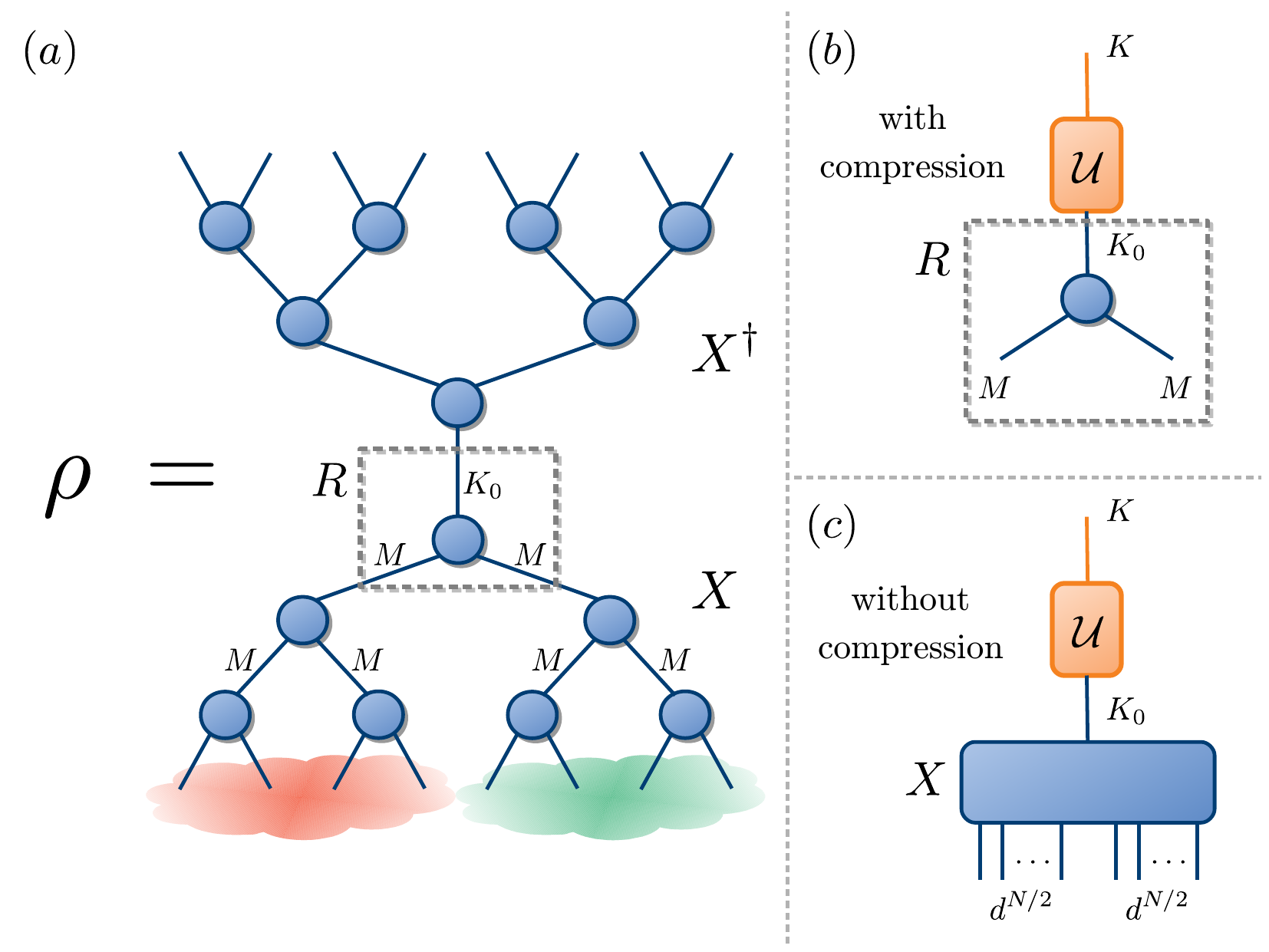}
	\caption{
	(a) The Tree Tensor Operator (TTO) representing a density matrix $\rho = X X^\dagger$. $K_0$ is the number of pure states in the representation used, while $M$ is the maximal  dimension for all bonds. 
	The gray dashed square highlights the root tensor $R$, containing all the information about entanglement between the red and green bipartitions of the physical space. 
	(b) Change of representation for the \eof minimization using $R$, after having compressed the state with some maximal bond dimension $\bd$. 
	(c) Same as (b), but without compression, so that $\bd = d^{N/2}$. 
	Optimizations are possible for any system size and state that can be efficiently represented as TTOs.
	}
	\label{fig:TTO}
\end{figure}
%
Conversely, for {\it mixed} global quantum states, the problem of characterizing and quantifying bipartite entanglement
is much more involved, both conceptually and technically. It is nevertheless a fundamental goal, since any realistic
quantum platform faces imperfections, statistical errors, and/or imperfect isolation leading to finite temperatures.
From a conceptual standpoint, a major focus is to assess which of the entanglement monotones proposed over the years satisfy
the desired properties of entanglement measures \cite{Bruss_JMP02}.
At a technical level, the core problem is to efficiently estimate these entanglement quantifiers. Even those that can be evaluated by linear algebra operations,
such as negativity \cite{Vidal_negativity} and quantitative witnesses \cite{Jens_QEW,Silvi_QEW}, are exponentially expensive in the system size.
Additionally, many important monotones with a clear physical significance, in terms of resource and information theory, are
{\it convex-roof} extensions of pure-state entanglement measures \cite{Plenio_ReviewEntanglement}.
Estimating these monotones is a hard non-linear minimization problem over pure-state decompositions of the global density matrix
\cite{Zyczkowski_PRA99,Audenaert_PRA01,Caro_PRA08,Allende_PRA15,Toth_PRL15,Loss_PRL08, Loss_PRA09}, severely limited to small system sizes.
%
%
%

%
The key point of our strategy is to exploit TN compressing capabilities and the exploitation of the \textit{Tree Tensor Operator (TTO)}  structure to represent a density matrix $\rho$ (Fig.~\ref{fig:TTO}) 
\cite{MPSReps,Uli_ageofMPS,VidalTTN2D,GersterTTN,SimonesBook,Silvi_Anthology19,
Evenbly_PRL15}.
This TN ansatz guarantees positivity of $\rho$, and being loopless it is efficiently contractible. Moreover, it is a natural TN geometry for estimating bipartite entanglement measures:
as discussed below, the information about bipartite entanglement is compressed into a single tensor,
ultimately simplifying the complexity of the minimization problem.
We demonstrate this method effectiveness computing the \eof of thermal many-body quantum states of the 1D transverse-field Ising and XXZ models.

{\it Tree Tensor Operator ansatz}
$-$ As positive operators, density matrices $\rho = \sum_j p_j \ketbra{\psi_j}{\psi_j}$
can be written as $\rho = X X^\dagger$, where the rectangular matrix $X = \sum_j \sqrt{p_j} \ketbra{\psi_j}{j}$
has a number of columns equal to the rank of $\rho$, also known as the the Kraus dimension $K_0$.
For many-body quantum states at low temperatures, probabilities $p_j$ decay sufficiently fast that it is possible
to approximate $\rho$ using a $K_0$ that scales at most {\it polynomially} with the system size $N$.
Therefore, from a numerical viewpoint, it is meaningful represent $X$ with a Tree Tensor network as shown in
Fig.~\ref{fig:TTO}: the  lower open links (`leaves', each of dimension $d$) represent the physical sites, while the upper open link (`root', of dimension $K_0$) represents the Kraus space of the global purification.
As for other Tensor Network ans{\"a}tze, this representation becomes efficient when the connecting links, or `branches', carry an effective
dimension $M$ that also scales polynomially with $N$ \cite{SimonesBook,Silvi_Anthology19,orus2019}.
%

By construction, the TTO ansatz \textit{guarantees positivity} of $\rho$, in contrast to the Matrix Product Density Operator ansatz~\cite{Verstraete_PRL04,Zwolak_PRL04}, whose positivity can be checked only as an NP-hard problem~\cite{Kliesch_PRL14}.
Locally Purified Tensor Networks~\cite{Werner_LPTN_PRL16} also preserve positivity, but the presence of loops in their network geometry leads to numerical limitations when implementing optimization strategies \cite{Gemma2013,Gemma2020}.
The TTO is instead positive and \textit{loopless} 
thus encompassing the best of the two words without any drawbacks.
When the TTO is properly isometrized to the {root tensor}, via (efficient) TN gauge transformations \cite{Silvi_Anthology19},
all the information about the mixing probabilities $p_j$ ends up stored within that tensor. Thus, also information
about global entropies (Von Neumann $\mathcal{S} = - \sum p_j \log p_j$ and R{\'e}nyi $\mathcal{S}_{\alpha} = (1-\alpha)^{-1} \log \sum_j p_j^{\alpha}$, including the purity). Moreover, all the information on bipartite entanglement (for a half-half system bipartition) is contained only in the root tensor. Indeed, the action of the isometrized branches is actually an {\it invertible LOCC}
(operation achievable via Local Operations and Classical Communication), and entanglement monotones cannot increase under such transformations \cite{Bruss_JMP02}.
In conclusion, compressing the relevant information into a tensor with polynomially-scaling dimension, it is possible to efficiently estimate entanglement monotones by processing only the root tensor, even for complex measures that rely on convex-roof extensions. Below, we specialize this procedure to the specific case of the EoF.

%
%
%
%
%
%

{\it \eof estimation $-$}
The \eof of a mixed quantum state $\rho$, defined as~\cite{Plenio_ReviewEntanglement} 
\begin{equation*}	\label{EOF:eq:EOF_definition1}
	E_{F}(\rho) = \inf_{\{p_j,\psi_j\}} \Big\{ \sum_j p_j \Svn(\ket{\psi_j}) : \rho = \sum_j p_j \ketbra{\psi_j}{\psi_j} \Big\}\,, 
\end{equation*}
quantifies the number of Bell pairs needed to construct a certain number of copies of $\rho$ via LOCC.
The minimization runs over all possible decompositions of $\rho$ as a convex mixture of pure states $\ket{\psi_n}$, with probabilities $p_n$.
%
It is straightforward to recast the previous expression in terms of the matrix $X$, whose columns $\sqrt{p_j} \ket{\psi_j}$ represent one possible pure-state decomposition of $\rho$.
Via the Schr\"odinger-HJW theorem~\cite{HJW_theorem,NielsenChuang}, it is possible to 
obtain the whole set of $X'$ matrices representing $\rho$, and thus all possible pure-state decompositions. This is done by multiplying
$X' = X \mathcal{U}$, where $\mathcal{U}$ is any right-isometry (a semi-unitary matrix satisfying $\mathcal{U} \mathcal{U}^{\dagger} = \mathbb{1}$)
of dimension $K_0 \times K$, with $K \geq K_0$. The minimization problem 
then becomes a minimization
over the space of right isometries $\mathcal{U}$, precisely
%
\begin{equation}	\label{EOF:eq:EOF_definition2}
	E_{F}(\rho) = \min_{K \geq K_0} \inf_{\mathcal{U}} \Big\{ \sum^{K}_{j=1} p_j \Svn(\ket{\psi'_j}) : X' = X \mathcal{U} \Big\} \; , 
\end{equation}
where the columns of $X'$ represent the new pure-state decomposition of $\rho$, with wavefunctions $\ket{\psi'_j} = X'\ket{j} (p'_j)^{-1/2}$
and probabilities $p'_j = \bra{j} X'^{\dagger} X' \ket{j}$. 

As depicted in Fig.~\ref{fig:TTO}(a), the $X$ matrix composing the isometrized TTO can be written as $X = (\mathcal{V}_L \otimes \mathcal{V}_R) R$, where $R$ is the root tensor, and the branches
$\mathcal{V}_{\star}$ are left-isometries ($\mathcal{V}_{\star}^{\dagger} \mathcal{V}_{\star} = \mathbb{1}$). It follows that the columns of $R$ must have the same
entanglement entropy $\mathcal{S}$ of the columns of $X$, and clearly the same probabilities $p_j'$. Thus, Eq.~\eqref{EOF:eq:EOF_definition2} can be more efficiently computed by replacing $X$ with the smaller root tensor $R$.

{\it Numerical Simulation} $-$ Hereafter, 
we estimate the EoF of low-temperature many-body states of 1D quantum lattice models $H$ via TTO. 
The first step is to approximate the many-body density matrix as a TTO ansatz. That is, writing
$X = \frac{1}{\sqrt{Z}} \sum_j^{K_0} e^{-E_j/2T} \ket{\psi_j} \bra{j}$
in tensor network format as it appears in Fig.~\ref{fig:TTO},
where $E_j$ is the energy of the Hamiltonian eigenstate $\ket{\psi_j}$ and $Z$ ensures normalization $\trace{X X^\dagger} = 1$. 
In this work, we achieve this goal with either of the two following methods:
{\it (i)} Energy eigenstates $\ket{\psi_j}$ are obtined via Exact Diagonalization (ED). Thus $X$ is calculated exactly, for a given $K_0$, and then
compressed into a TTO as detailed in the Supplemental Material (SM).
{\it (ii)} $\ket{\psi_j}$ are obtained via a Tree-Tensor Network algorithm capable of targeting each of the $K_0$ lowest energy eigenstates \cite{GersterHofstadter}. Their collected information
is then easily formatted into a TTO with standard tensor network operations \cite{Silvi_Anthology19}.
The accuracy of this second approximate method is benchmarked against the exact one for thermal states of small size in the SM. 
Beyond these two strategies, we envision the possibility to develop algorithms that directly compute the TTO for finite-temperature quantum states, capture Markovian real-time evolution, or transform other TN states into TTOs~\cite{Inprep, Orus_PRL19}.

Once the TTO is built, we proceed to calculate the optimization from Eq.~\eqref{EOF:eq:EOF_definition2} on the top tensor $R$.
To build sets of $\mathcal{U}$ matrices, we fix a value for $\nstates \geq \nstates_0$ 
and parameterize a Hermitian matrix $A = A^{\dagger}$ of dimensions $\nstates \times \nstates$. Then, we get the corresponding unitary from 
$U = \exp\{iA\}$, and finally we take $K_0$ random rows of $U$ to build $\mathcal{U}$.
For every column of $R' = R \mathcal{U}$, its entanglement entropy is calculated via $\Svn = -\sum_i s_i^2 \log s_i^2$,
where the singular values $s_i$ are obtained by a singular value decomposition (SVD). In the results section, entropies
are expressed in basis of $\log_2$, so that a Bell pair defines the unit of entanglement.
For a given $K \geq K_0$, minimization in the space of the $\mathcal{U}$ is carried out via direct search methods, but other choices are possible. 
Extensive proofs of the stability of this method, as well as some results on many-body random density matrices, are provided in the SM. 
Convergence of the minima is rapidly reached when increasing $K \geq K_0$. For all practical purposes, choosing $K \approx K_0$ is often sufficient to achieve close convergence (see SM). 
We stress that, even in case of incomplete or failed convergence, our method still provides an upper bound to the actual \eof of the quantum state. In particular, in every case we could check, the results provided tight bounds. 
The accuracy and convergence of the \eof estimation is discussed in detail in the SM: 
we $(i)$ benchmark the optimization procedure applied to exact states of small systems, whose \eof is known a priori, and 
$(ii)$ test how state compression into a TTO affects the \eof computation, by comparing optimizations done on the approximate root tensor $R$ and on the exact $X$, for thermal states of small systems.

{\it Results} $-$
%
We consider two well-known prototype quantum critical spin-$\frac{1}{2}$ models as benchmarks~\cite{Takahashi_book99}: 
specifically, the Ising model
\begin{equation}		\label{Res:eq:Ising}
	\Hop_{\text{Ising}} = J \sum_{i=1}^N \left( 
		\sx_j \sx_{j+1} + h \sz_j 
	\right)
\end{equation}
in a transverse field $h$, and the XXZ model
\begin{equation}		\label{Res:eq:XXZ}
	\Hop_{\text{XXZ}} = J \sum_{j=1}^N \left(
		\sx_j \sx_{j+1} + \sy_j \sy_{j+1} + \xi\, \sz_j \sz_{j+1} 
	\right)
\end{equation}
with anisotropy $\xi$, both models considered in periodic boundary conditions (PBC) and $\hat \sigma_j^\alpha$s ($\alpha=x,y,z$) are the Pauli matrices.
The temperature $T$, defining the thermal state $\rho = \frac{1}{Z} e^{-\Hop/T}$, is expressed in units of the Hamiltonian energyscale ($J= k_B = 1$).
To appropriately choose a suitable number $K_0$ 
we start from $K_0 = 2$. We then evaluate the resulting \eoff, gradually increasing $K_0$ until convergence of the estimated EoF is reached. We employ a similar strategy to choose the best $\bd$. 
\begin{figure}
	\centering
	\includegraphics[width=\columnwidth]{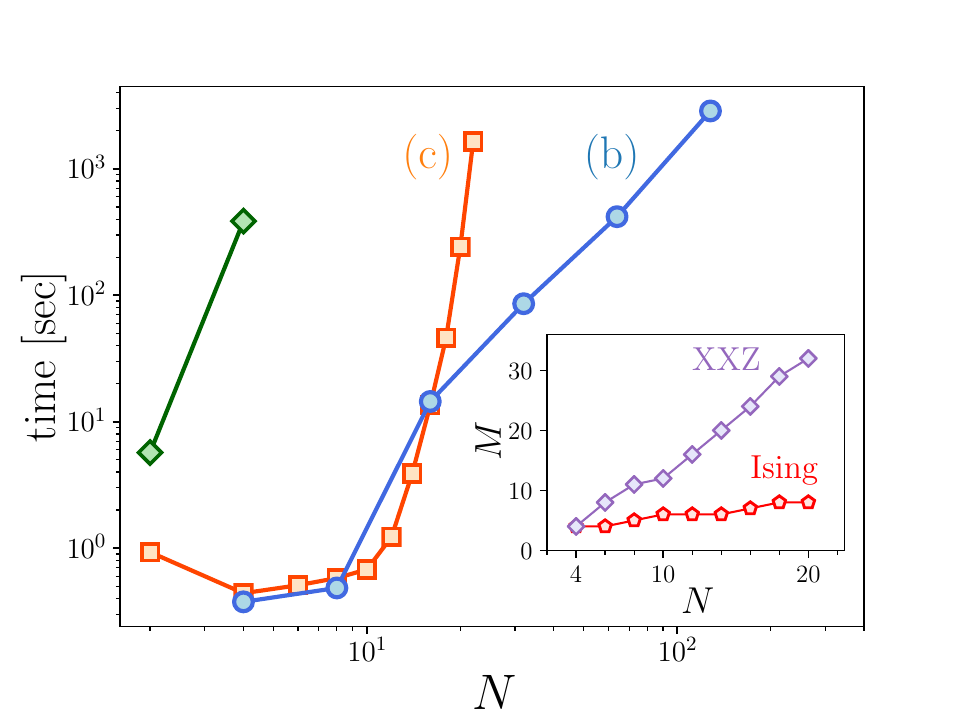}
	\caption{
	Computational times of EoF estimation `from scratch' as a function of system size $N$, for critical $\Hop_{\text{Ising}}$ ($h=1$) at temperature $k_B T = 0.5 \Delta$, where $\Delta \sim N^{-z}$ ($z=1$) is the finite-size gap.
	Two data sets show EoF Estimation without TTO compression:
	either using the full exact density matrix (green diamonds), or from the exact $X$ matrix at convergence in columns $K_0$ (orange squares).
  The last data set shows EoF estimation with TTO method, where the Hamiltonian eigenstates were calculated via Tree-Tensor Network eigensolver algorithm.
	\textit{Inset:} 
	Bond dimensions $\bd$ required to achieve convergence of the \eof estimator (approximation less than $1\%$ from its exact value).
	Red pentagons and purple diamonds refer respectively to 
	the critical Ising model at $k_B T = 0.1 J$ and to 
	the XXZ model with $\xi = 0.5$ (critical) at $k_B T = 0.5 J$. 
	}
	\label{Res:fig:time_scaling}
\end{figure}
Fig.~\ref{Res:fig:time_scaling} shows a typical benchmark comparison of the total computational time required to acquire the \eoff, which include both calculation of the thermal state and entanglement measure estimation. Three data sets show respectively \eof estimation using:
the full exact density matrix (green),
the exact $X$ matrix at convergence in $K_0$ (red), and
the TTO method with Tree-Tensor Network eigensolver (blue).
Complexity of the exact methods increases exponentially, basically as $\osmall( {\dimH}^{3/2} )$, since the bottleneck of our algorithm is the SVD to calculate $\Svn$ for each of the $K$ pure states.
By contrast, this runtime scales like $\osmall( M^3 )$ for a TTO representation, with $\bd \ll \sqrt{\dimH}$.
At size 20 and beyond, the TTO algorithm clearly outperforms exact methods, displaying a visibly polynomial scaling.
Typical bond dimensions $M$ required to have a good approximate estimation ($99\%$ of the exact \eof value) are displayed in the inset, and show a roughly linear scaling with the system size.
Equipped with our diagnostic tool, we quantified the bipartite entanglement properties of two
quantum systems at finite $T$. 
The two panels in Fig.~\ref{Res:fig:scale_invariance} focus on critical phases of the two models,
the quantum phase transition point of the Ising model ($h=1$, top), and the Luttinger liquid phase of the XXZ model ($\xi = 0.5$, bottom) respectively.
While the system is strongly-correlated at zero temperature, entanglement seems to survive roughly unaltered up to $T$ 
of the order of $0.2 \Delta(N)$, 
with $\Delta(N)$ the finite-size energy gap, 
and smoothly drop at higher $T$. 
This phenomenon is to be contrasted with the Von Neumann entropy $\mathcal{S}$  (global, or of either subsystem), which instead {\it grows} with $T$, and can not capture alone
the entanglement decrease~\cite{Calabrese_JSTAT04,Calabrese_JPA09}. 
More importantly, we observe an emergent scaling behavior when plotting $E_F(T,N)$. In fact, the \eof appears to follow the logarithm of a conformal scaling function,
in proximity of the quantum critical point (i.e.,~for small temperatures $T \sim \Delta$). For PBC, this behaviour can be expressed as
$E_{F} = \log ( N^{c/3} f( T N^{z} ) )$, or
\begin{equation} \label{Res:eq:invariance}
 E_{F} (T,N) = \frac{c}{3} \log N + g( T N^{z} )
\end{equation}
in analogy to Ref.~\cite{FisherBarber}, where $c$ is the critical exponent that connects
lengthscales to entanglement, while $z$ is the critical exponent that connects lengthscales to energyscales ($\Delta \propto N^{-z}$). The functions $f(\cdot)$ and $g(\cdot) = \log f(\cdot)$ are  non-universal and depend on the microscopical details of the model.
This behaviour actually extends, to finite $T$, the known
scaling law for the entanglement entropy with size, valid for critical ground states \cite{Calabrese_JSTAT04,Calabrese_JPA09}.
We validate this argument in the inset of Fig.~\ref{Res:fig:scale_invariance}, where the $E_F(T,N)$ data sets are appropriately rescaled, according to $N$.
As we expect, the curves collapse when the appropriate critical exponents of the corresponding model are used
($c = \frac{1}{2}$, $z=1$ for critical Ising; $c = 1$, $z=1$ for Luttinger liquid XXZ). In the former case, we pushed the numerics to very large system sizes and fully exploited the TTO approach, while in the latter we limited our analysis to ED methods as it clearly suffices to confirm the scaling behaviour of critical systems.

\begin{figure}
	\centering
	\includegraphics[width=\columnwidth]{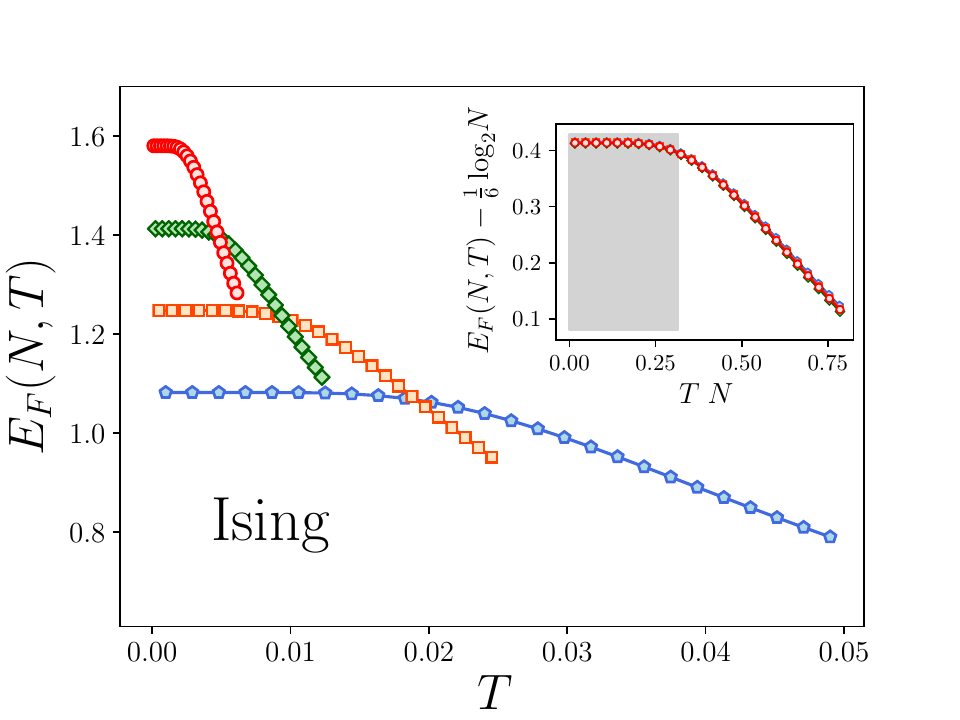}
	\includegraphics[width=\columnwidth]{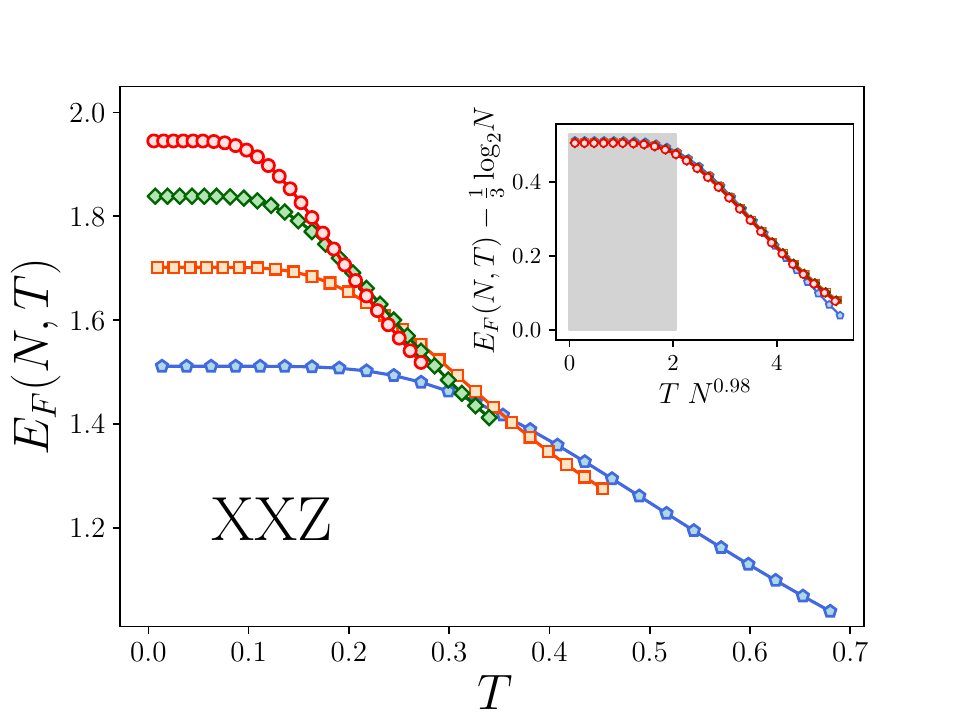}
	\caption{Scale-invariance of the \eof at temperatures $T$ 
	(in units of $J/k_B$) in the range 
	$k_B T \leq 0.5\Delta$, where $\Delta \propto N^{-z}$, 
	for the critical Ising model in Eq.~\eqref{Res:eq:Ising} (top panel) and the XXZ model in Eq.~\eqref{Res:eq:XXZ} in the critical phase at $\xi=0.5$ (bottom panel). 
	Data for $N = 16, 32, 64, 128$ (top, approximate method for TTO computation) and $N = 8, 12, 16, 20$ (bottom, exact TTO from ED), from the flattest to the steepest curve. 
	\textit{Inset:} curves in the main figures after rescaling according to Eq.~\eqref{Res:eq:invariance}.
	The agreement is stunning, using $c=1/2$ and $z=1.00 \pm 0.01$ (top) 
	and $c=1$ and $z=0.98 \pm 0.02$ (bottom). 
	The grey area highlights the temperature range $T \leq 0.2 \Delta(N)$. 
	}
	\label{Res:fig:scale_invariance}
\end{figure}

As a final remark, we stress that the \eof analysis enabled by the TTO method is not limited to low-temperature many-body states of lattice models.
We have employed the same diagnostic tool on other classes of mixed many-body states, including on sets where the \eof is known to further benchmark our approach, as reported in the SM.

\paragraph*{Conclusions}
In this letter, we have presented a new tensor network approach that enables the numerical analysis of bipartite entanglement for many-body quantum systems, even for
those entanglement monotones that are considered {\it hard} since they require convex-roof optimization.
We employed a Tree Tensor Operator (TTO) to well-approximate the global density matrix at low temperatures. Such a tensor network architecture compresses information
of the bipartite entanglement into a single tensor, whose dimensions in many cases scale polinomially with the system size. As a result, evaluating entanglement monotones is numerically efficient, as illustrated for 1D interacting lattice models.
Our analysis observed a scaling law for the Entanglement of Formation, compatible with a logarithmic conformal scaling law.
We successfully tested this argument for a free fermion (Ising) and an interacting fermion (XXZ) critical models, where it is satisfied in a temperature range commensurate with the finite-size energy gap ($T \sim \Delta$).
%

While we built TTOs by collecting information on low-lying energy eigenstates, we envision the possibility of developing algorithms capable of directly targeting finite-temperature states on a TTO architecture.
Similarly, we envision the possibility of replacing the TTN branches of the ansatz with Matrix Product State branches: an alternative TN design that is still efficient toward \eof estimation.
Finally, we expect that TTO may be capable to accurately capture some features of open-system quantum dynamics. This will actually extend the bipartite-entanglement analysis, presented here, from finite-temperature states to a larger set of open-system physically relevant states, i.e. the stationary states of a Lindblad master equation \cite{Schaller_book, Cui_PRL15,Savona_PRA15}. The Time-Dependent Variational Principle~\citep{TDVP_PRL11,TDVP_PRB16} is surely a good candidate
strategy towards this goal. This will likely be the focus of our research in the near future.
%
%

\acknowledgements

We thank M. Gerster for his support in software development,
as well as M. Dalmonte and B. Kraus for stimulating discussions.
Authors kindly acknowledge support from 
the Italian PRIN2017 and Fondazione CARIPARO, the Horizon 2020 re-search and innovation programme under grant agreementNo 817482 (Quantum Flagship - PASQuanS), the Quan-tERA projects QTFLAG and QuantHEP, the DFGproject TWITTER, 
and 
the Austrian Research Promotion Agency (FFG) via QFTE project AutomatiQ. 
We acknowledge computational re-sources by the Cloud Veneto. 
This research was supported in part by the National Science Foundation under Grant No. NSF PHY-1748958.

\bibliographystyle{apsrev4-2}
\bibliography{biblioTTO}

\cleardoublepage
\section{Supplemental material}

\subsection{Algorithm and its computational cost}
In this section, we illustrate the basic operations done at each minimization step, to obtain the computational cost of the algorithm. 
Given a $K \times K$ hermitian generator $A$, whose entries are the parameters to optimize, the corresponding $\mathcal{U}$ (of dimension $K_0 \times K$)
is constructed by keeping the top $K_0$ rows of
$e^{i A}$, which in turn is computed by matrix exponentiation.
Such computational cost is roughly $\osmall(K^3)$, theoretically improvable to $\osmall(K^\omega)$ with $2 < \omega < 3$ with fast matrix-multiplication methods \cite{demmel2007fast}.
Then, we contract $\mathcal{U}$ with the initial state tensor --- be it $X$ or $R$ --- along the Kraus link, which yields a new pure-state decomposition.
This operation costs $\osmall(d^N K_0 K)$ for the full picture $X$, but only $\osmall(M^2 K_0 K)$ when using the TTO method, i.e.~using $R$.

To compute the figure of merit, which we are minimizing according to Eq.~(1) in the main text, we need the probability and
entanglement entropy for each of the $K$ states in the new ansatz ensemble.
While acquiring the probabilities has subleading cost, extracting $\Svn$ is carried out by repeating $K$ times a Singular Value Decomposition (SVD),
which contributes to the overall cost with $\osmall(K d^{3N/2})$ for the full $X$, while only $\osmall(K M^3)$ for the compressed $R$ of the TTO method.
For the low temperatures considered in this work, we could safely work under the assumption $K \lesssim M$, which in turn makes the latter contribution
the leading computational cost.

These estimated computational costs match the observed scaling of runtime that we reported in Fig.~2 in the main text.
%
%

\subsection{Root tensor extraction from a density matrix}
\begin{figure}
	\centering
	\includegraphics[width=\columnwidth]{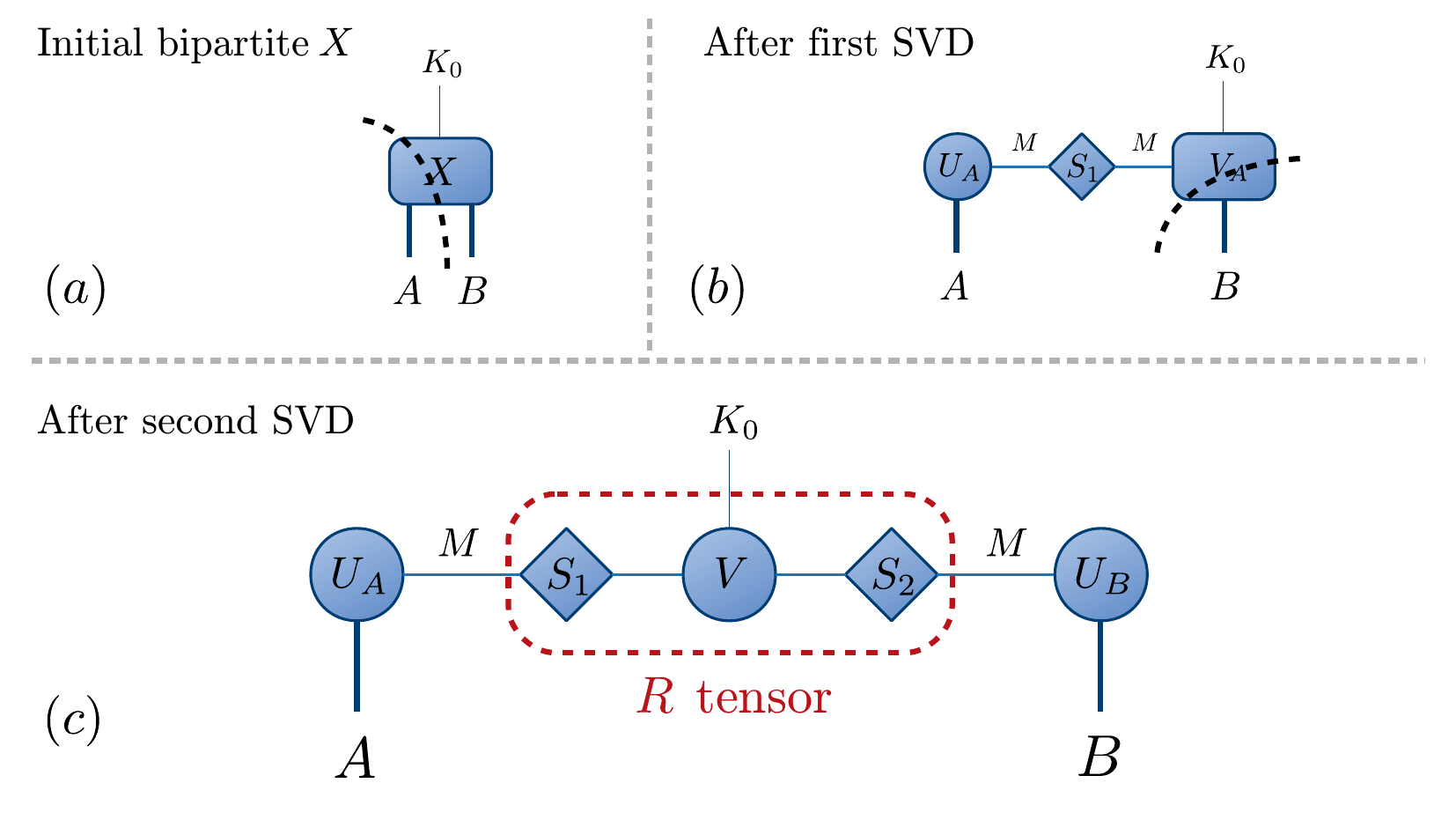}
	\caption{How to obtain the TTO root tensor $R$ from the full density matrix representation $X$. 
	The black dashed lines in panel~(a) and (b) show how to fuse legs before doing the SVD. 
	After the second SVD, the root tensor $R$ is just the contraction of the three tensors within the red dashed curve in panel (c). Notice the truncation to the $\bd$ largest singular values. }
	\label{SM:fig:TTOsplitting}
\end{figure}

Given a density matrix represented by a matrix $X$ as specified in the main text, 
Fig.~\ref{SM:fig:TTOsplitting} illustrates how to obtain the root tensor $R$ of its corresponding TTO. 
First, the physical leg of $X$ is split in two smaller legs according to the system bipartition, the left (right) leg representing the physical space of bipartition $A$ $(B)$ (see panel~(a)). 
We then fuse the legs to the right of the black dashed line, obtaining a matrix again. 
Then, a Singular Value Decomposition (SVD) with truncation up to the $\bd$ largest singular values is performed: 
this provides what can be seen in panel~(b), with an isometry $U_A$ linked to one of the two bipartitions, a diagonal matrix $S_1$ with the $\bd$ largest singular values, and an isometry $V_A$ which contains also the Kraus leg. 
Then, we fuse the links of $V_A$ that lie above the black dashed line in panel~(b) and perform on it another SVD with truncation, to obtain the outcome in panel~(c). 
At this point, there are two isometries $U_A$ and $U_B$ linked to their own physical bipartitions, and information about entanglement between them is all contained in the root tensor $R$ coming from contraction of all the remaining tensors (the ones included within the red dashed line in panel~(c)). 
Notice that $R$ has the two lower legs upper bounded by the maximal bond dimension $\bd$. 
The overall computational cost of this procedure is roughly $\osmall(\bd^3 K_0)$. 

\subsection{On the choice of $K_0$ and $K$}
Let us first illustrate the strategy to choose the Kraus dimension $K_0$ for the initial representation of thermal states
$\rho = \frac{1}{Z} \sum_j e^{-E_j/k_B T} \ket{\psi_j} \bra{\psi_j}$.
%
If the temperature $T$ is relatively small, the weights of the excited states $\sim e^{-E_j/k_B T}$ decay rapidly to zero. Even for critical systems, where the excitation energy-scale
is only given by the finite-size gap $\Delta$ and decays with $N$ as a power-law ($\Delta \propto N^{-z}$), 1D excitations are typically exponentially suppressed. 
%
To find how many eigenstates ($K_0$) are necessary to well-approximate the thermal state, we adopt a verification strategy based on the \eof.
Precisely, we compute the \eof of mixed states obtained from the thermal one after truncation to the $K_0$ eigenstates with largest probabilities. 
We thus check the scaling of the \eof for increasing $K_0$ and locate when it reaches convergence. 
%
In doing so, we always fix $K = K_0$. 
At the end of the optimization for a given $K=K_0$, the optimal solution is used as the starting point for the next optimization with $K_0+1$ eigenstates. 
We do this by simply adding a row and a column of zeros at the bottom and to the right of the Hermitean matrix parameterizing $\mathcal{U}$. 
%
\begin{figure}
	\centering
	\includegraphics[width=\columnwidth]{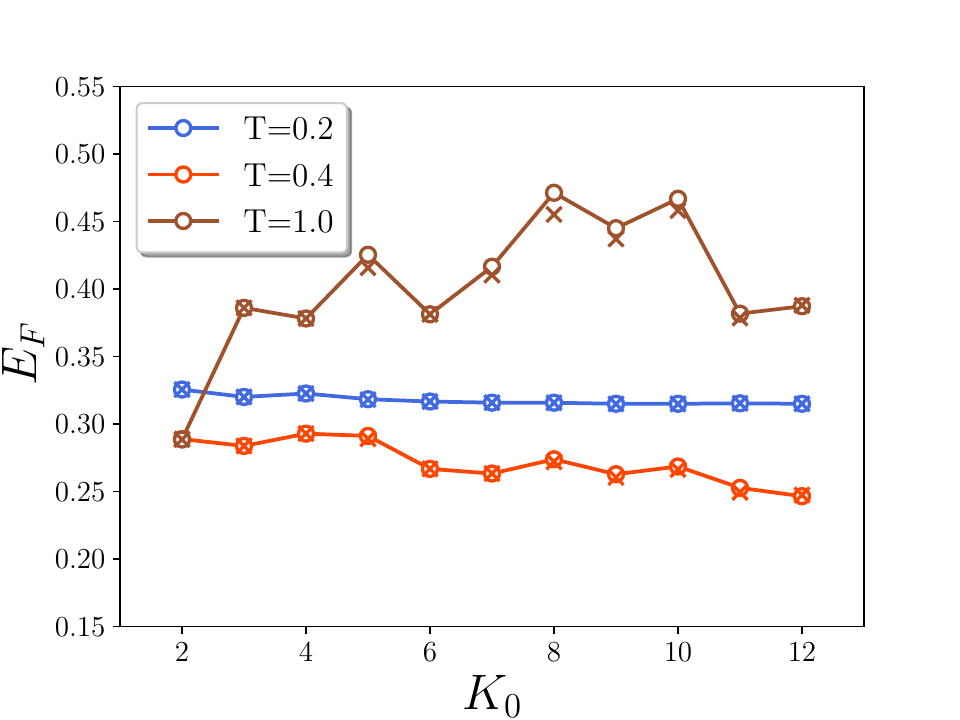}
	\caption{Scaling of $E_F$ for thermal states of the critical Ising model in transverse field ($N=16$) at different temperatures \textit{versus} the number of states $K_0$ kept in the thermal ensemble. 
	Temperatures in the key are expressed in units of $J/k_B$. 
	Circles correspond to data for $K = K_0$, while crosses are for $K = K_0 + 2$. 
	From the results, it is clear that $K = K_0$ is already enough to achieve a very good estimate for the \eof.
	 } 
	\label{SM:fig:K0scaling}
\end{figure}

Fig.~\ref{SM:fig:K0scaling} shows results for the critical Ising model in transverse field with $N=16$ sites and at different temperatures $T$. 
The \eof for $k_B T \leq 0.2 J$ converges very fast already with $K_0 = 2$, while larger temperatures require more states. 
%
Circles refer to optimizations with $K = K_0$, while crosses correspond to $K = K_0 + 2$. 
We observe that there is no need to enlarge the parameter space for this class of problems, since each circle is superimposed or very close to its corresponding cross. 

\begin{figure}
	\centering
	\includegraphics[width=\columnwidth]{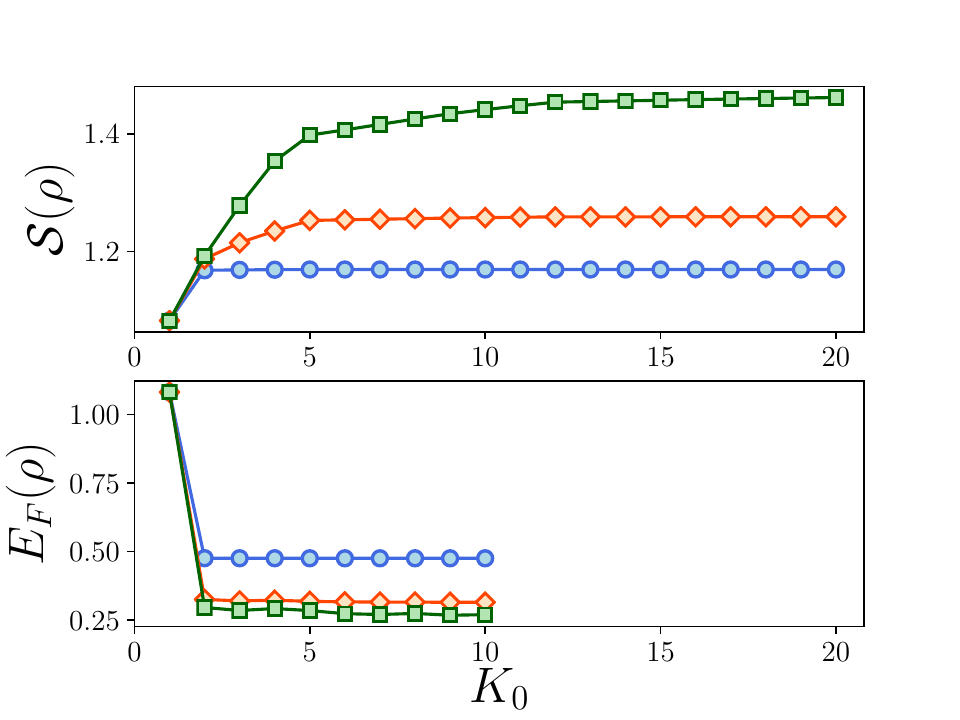}
	\caption{Convergence of the Von-Neumann entropy $\Svn(\rho)$ (top) and of the \eof $E_F(\rho)$ (bottom) in the number $K_0$ of states considered in the thermal ensemble. 
	Data refer to the critical Ising model with $N=16$ spins at temperatures $k_B T=0.1,0.2,0.3 J$ (blue circles, orange diamonds and green squares, respectively).
	We observe that $E_F(\rho)$ estimation requires less states with respect to $\Svn(\rho)$.}
	\label{SM:fig:VNE_vs_EOF}
\end{figure}

One might be tempted to pinpoint the best $K_0$ by looking at other quantities, possibly easier to calculate. However, we find this can be deceptive: we support this statement by looking at the Von-Neumann entropy of the density matrix, $\Svn(\rho)$, instead of its \eof $E_F(\rho)$. 
Fig.~\ref{SM:fig:VNE_vs_EOF} shows how the two quantities both converge for thermal states of the critical Ising model with $N=16$ spins and at different temperatures. 
At temperatures $k_B T=0.2,0.3 J$, $\Svn(\rho)$ needs a higher number $K_0$ of states to represent the correct result with respect to $E_F(\rho)$. 
Therefore, we preferred to look at the \eof scaling rather than other observables, although the computational effort is much greater. 

\subsection{On the choice of $\bd$}
\begin{figure}
	\centering
	\includegraphics[width=\columnwidth]{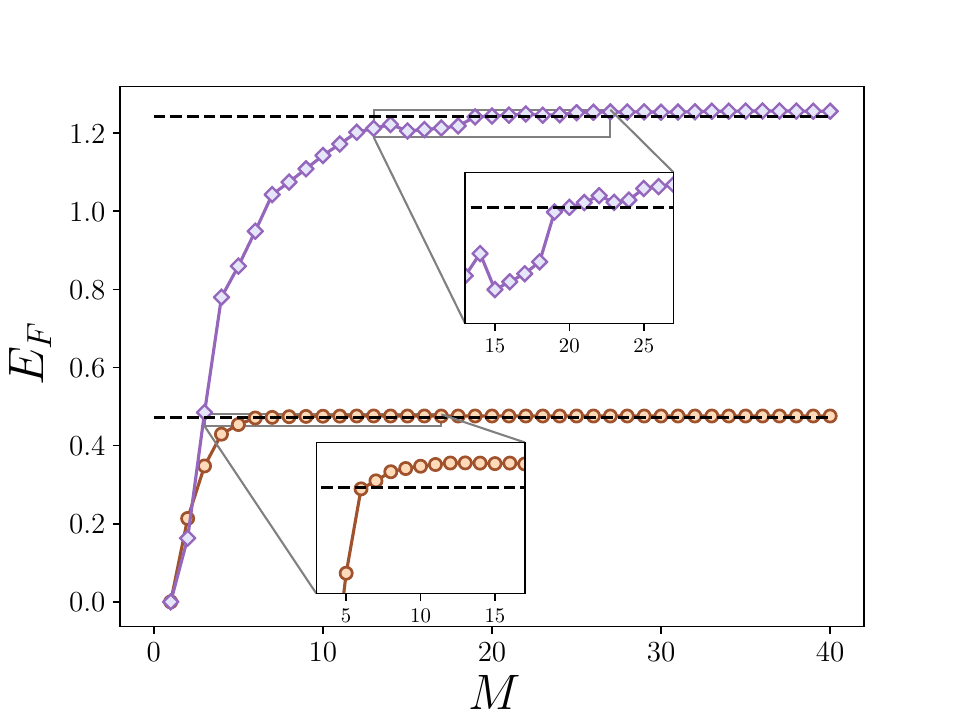}
	\caption{Scaling of $E_F$ for thermal states represented as TTO with different maximal bond dimensions $\bd$. 
	Brown circles and purple diamonds refer to the $N=16$ critical Ising model ($k_B T=0.1 J$) and XXZ model ($\xi=0.5$, $k_B T=0.5 J$), respectively. 
	The two black dashed lines show $99\%$ of the $E_F$ value at convergence. The insets zoom in the main plot, to help locating the smallest maximal bond dimension that represents the states well enough. 
	 }
	\label{Res:fig:Mscaling}
\end{figure}

\begin{figure}
	\centering
	\includegraphics[width=\columnwidth]{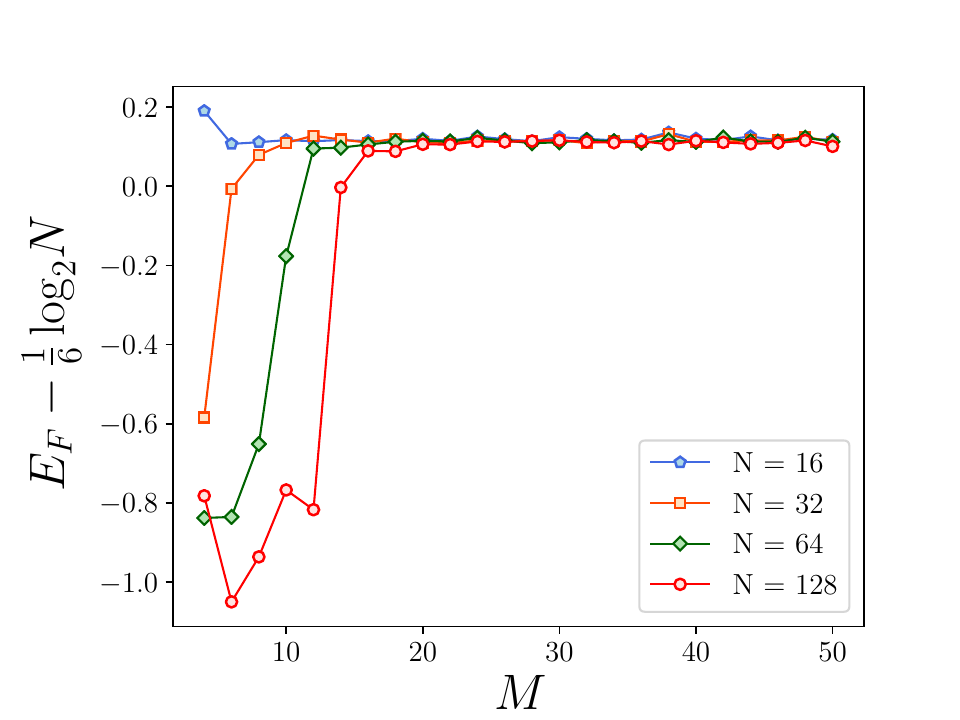}
	\caption{
	Scaling of $g(TN^z)$ in Eq.~(4) in the main text for thermal states of the critical Ising model represented as TTO with different maximal bond dimensions $\bd$, at $T = 0.5\Delta(N)$. 
	The TTOs are computed through a Tree-Tensor Network algorithm able to target the lowest $K_0$ energy eigenstates~\cite{GersterHofstadter}. 
	 }
	\label{Res:fig:Mscaling_TTO}
\end{figure}

Due to its structure, the TTO is exact whenever $\bd = \sqrt{\dimH}$. Therefore, to find the smallest maximal bond dimension $\bd$ to represent the state correctly, we plot the \eof $E_F$ for increasing $\bd$ starting from very low values and looking at when it reaches the $\bd \to \infty$ converged value within $1\%$. 
This is shown in Fig.~\ref{Res:fig:Mscaling} for two low-temperature thermal states of $N=16$ spins, computed through ED. Brown circles refer to the critical Ising model at temperature $k_B T=0.1 J$, while purple diamonds correspond to the critical XXZ model with $\xi = 0.5$ at $k_B T=0.5 J$. 
The black dashed lines point at $99\%$ of the converged $\bd \to \infty$ values: the first point for which all the subsequent ones are above this line corresponds to the smallest maximal bond dimension $\bd$. 
Notice that this is the criterion used to determine each data point in the inset of Fig.~2 in the main text. 
A similar calculation for TTOs computed via the Tree-Tensor Network algorithm for low-lying excited states~\cite{GersterHofstadter} is shown in Fig.~\ref{Res:fig:Mscaling_TTO}, where we plot the scale-invariant function $g(TN^z)$ in Eq.~(4) in the main text. 
With this approach, we reach sizes of hundred of sites, way beyond the capabilities of exact diagonalization.
We observed that this approach sometimes encounters numerical instabilities, requiring the repetition of the TTO calculation from three to eight times, roughly. This is due to the TTN algorithm for excited states sometimes failing to resolve a quasi-degeneracy in the spectrum, which in turn may lead to large fluctuations of the EoF for low (but non-zero) temperatures.
In this sense, EoF can also be used as a post-processing check of successful convergence.
%

\begin{figure}
	\centering
	\includegraphics[width=\columnwidth]{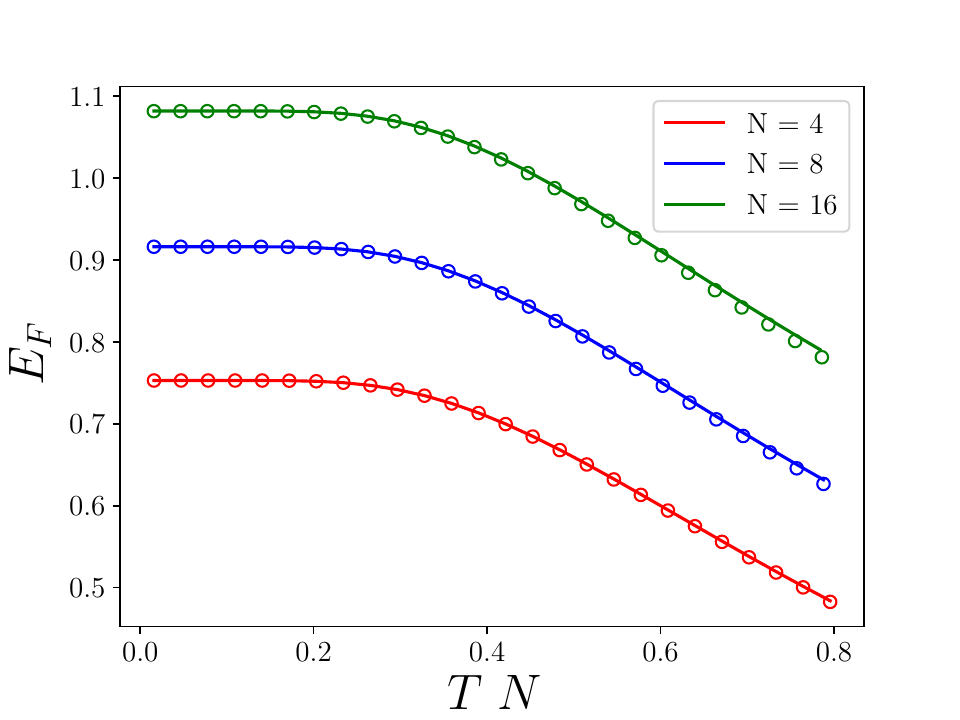}
	\caption{
	Benchmark of the approximated (circles) \vs exact (lines) approach for computing low-lying excited states, for the critical Ising model at small sizes $N = 4,8,16$ and at different temperatures. 
	Notice that temperatures are multiplied by $N$ for better illustration purposes, and if we did the \eof rescaling as in Fig.~(3) in the main text, the three curves would overlap. 
	 }
	\label{Res:fig:BenchmarkTTO}
\end{figure}

We end this section by illustrating the accuracy of the Tree-Tensor Network algorithm for low-lying excited states~\cite{GersterHofstadter}, as compared to results from exact diagonalization, for systems of small size. 
Fig.~\ref{Res:fig:BenchmarkTTO} shows excellent agreement between the \eof resulting from the exact (lines) and the approximate (circles) method for computing low-lying excited states of the critical Ising model. 
Notice that the three curves for different sizes would overlap if the rescaling in the insets of Fig.~3 was applied. 

\subsection{Benchmarks on non-thermal states}

We illustrate here several benchmarks done for the \eof to assess the reliability of our optimizations. 
We recall that the \eof for 2-qubits systems can always be computed exactly by using the Concurrence~\cite{Wootters_concurrence_PRL98}. 
We use this tool to study mixture of Bell and GHZ states, as well as some classes of random states. 
In the latter case, we also provide some results for larger numbers of qubits. 
We furthermore study Werner~\cite{Werner_PRA89,Werner_PRA01} and isotropic states~\cite{Isotropic_PRA99}, for which exact solutions exist for arbitrary Hilbert space dimension. 
Finally, we characterize the ability of our algorithm to detect zero entanglement in random separable states. 

\subsubsection{Bell and GHZ states}
\begin{figure}
	\centering
	\includegraphics[width=\columnwidth]{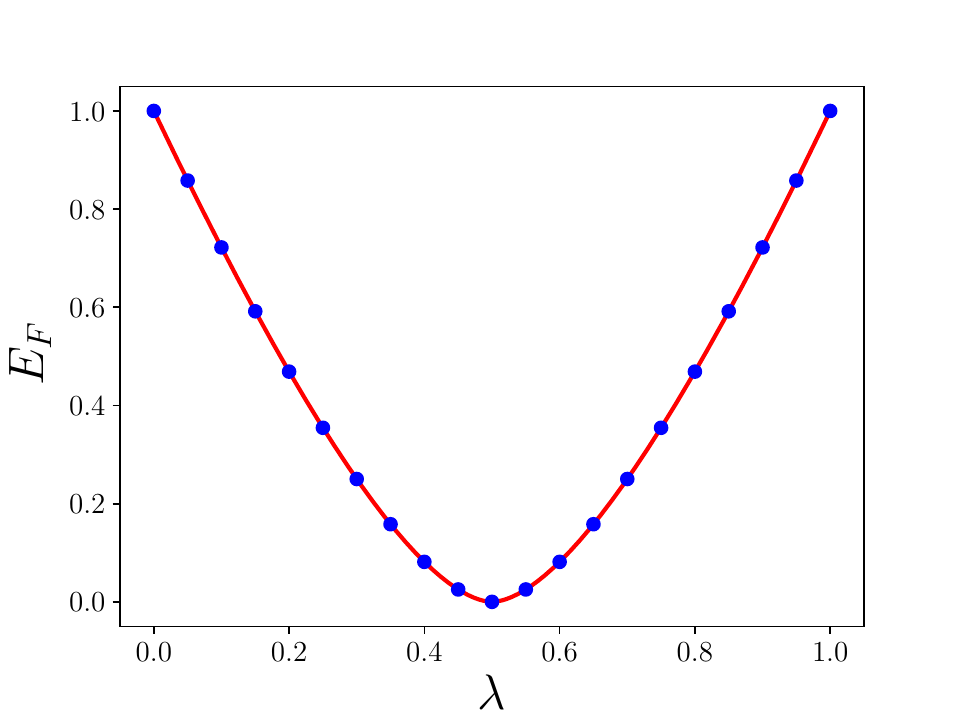}
	\caption{ $E_F(\rho)$ for $\rho$ from Eq.~\eqref{SM:eq:Bell}, for different values of $\lambda$. 
	The red line refers to exact results, while blue dots come from numerical optimization with $\nstates = \nstates_0 = 2$. 
	}
	\label{SM:fig:Bell}
\end{figure}

Consider the following mixture of two \textit{Bell states}: 
\begin{equation}	\label{SM:eq:Bell}
	\rho = 
	\lambda \ketbra{\phi_+}{\phi_+} + 
	(1-\lambda) \ketbra{\phi_-}{\phi_-} 
\end{equation}
where $\ket{\phi_\pm} = \left( \ket{\uu} \pm \ket{\dd} \right) / \sqrt{2}$ are the two Bell states taken into account and $\lambda \in [0,1]$. 
As a 2-qubits system, the exact \eof can be computed exactly~\cite{Wootters_concurrence_PRL98}. 
We thus compute the \eof via numerical optimization for some values of $\lambda$ and benchmark them against the exact results. 
This comparison is provided in Fig.~\ref{SM:fig:Bell}, showing excellent agreement already for $\nstates = \nstates_0 = 2$. 

The very same results hold also for mixtures of \textit{GHZ states} of $N$ qubits, 
that is Eq.~\eqref{SM:eq:Bell} where, instead of the Bell states $\ket{\phi_\pm}$, one takes the GHZ states 
$\ket{\GHZ_\pm} = \left( 
\ket{\uparrow \uparrow \dots \uparrow} \pm
\ket{\downarrow \downarrow \dots \downarrow}
\right) / \sqrt{N}$. 
Indeed, given a bipartition between subsystems $A$ and $B$, one can always rewrite the GHZ states as 
\begin{align}
\begin{split}
	&\ket{\uparrow \uparrow \dots \uparrow} = 
	\ket{\uparrow \dots \uparrow}_A \otimes \ket{\uparrow \dots \uparrow}_B \equiv 
	\ket{\Uparrow_A \Uparrow_B}	\; , \\
	&\ket{\downarrow \downarrow \dots \downarrow } = 
	\ket{\downarrow \dots \downarrow}_A \otimes \ket{\downarrow \dots \downarrow}_B \equiv 
	\ket{\Downarrow_A \Downarrow_B}	\; , 
\end{split}
\end{align}
so that entanglement is effectively the same as the one for the 2-qubits system in Eq.~\eqref{SM:eq:Bell}. 
It is insightful to see this in the TTO framework. 
Eq.~\eqref{SM:eq:Bell} for Bell states would correspond to a trivial TTO with a single (rank-3) tensor. 
If one takes $N$-qubits GHZ states instead of Bell states, the TTO would have more tensors, but its root tensor (the one with the Kraus link) would be identical to the trivial TTO for Bell states. 
As discussed in the main text, information about entanglement across a given bipartition is \textit{all} contained in the root tensor, so that the two states have indeed the same entanglement.

\subsubsection{Ensembles of random pure states with identical probabilities}
We study here $N$-qubits mixed states of the form 
\begin{equation}
	\rho = \sum_{n=1}^{K_0} \frac{1}{K_0} \ketbra{\psi_n}{\psi_n} \; ,
\end{equation}
where the pure states $\ket{\psi_n}$ are chosen at random according to a \textit{uniform} probability distribution over the set of all pure states of $N$ qubits. 
This is done by selecting randomly a unitary transformation according to the Haar measure~\cite{Mezzadri_RandomUnitaries_AMS07} and applying it to a reference state. 
Repeating this $\nstates_0$ times yields $\nstates_0$ pure states for the ensemble. 
Notice that all the probabilities associated to these pure states are chosen to be identical, for simplicity. 

\begin{figure}
	\centering
	\includegraphics[width=\columnwidth]{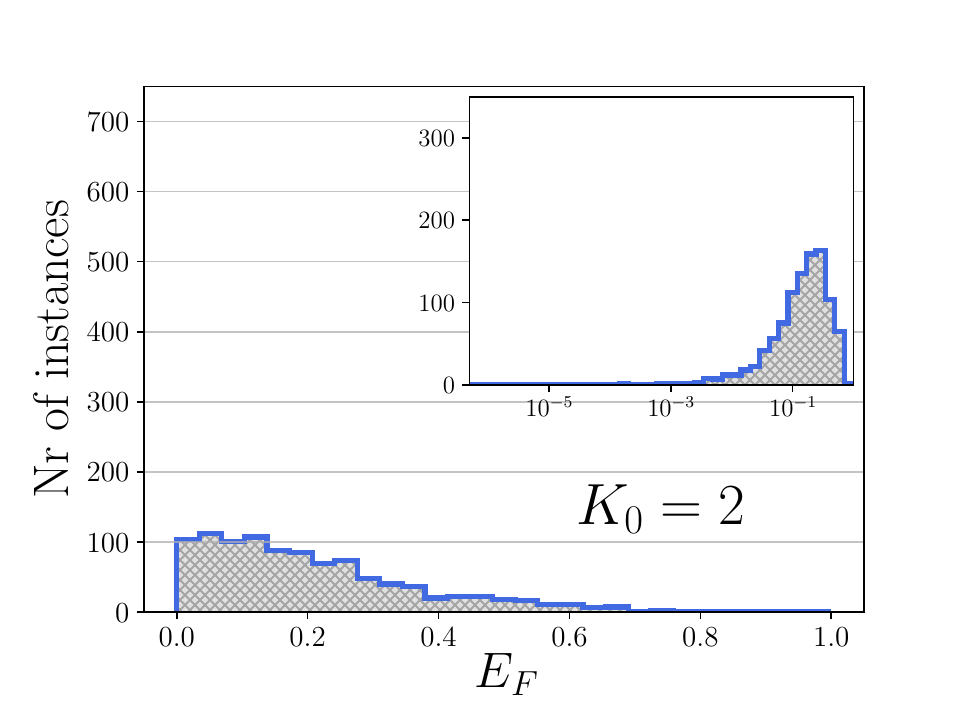} \\
	\includegraphics[width=\columnwidth]{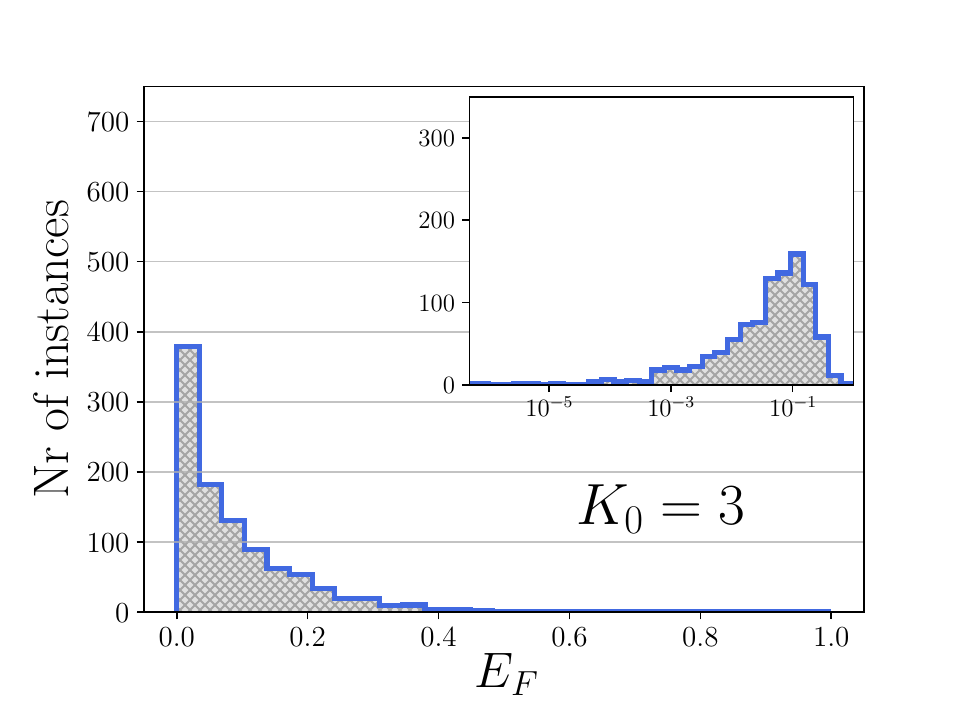} \\
	\includegraphics[width=\columnwidth]{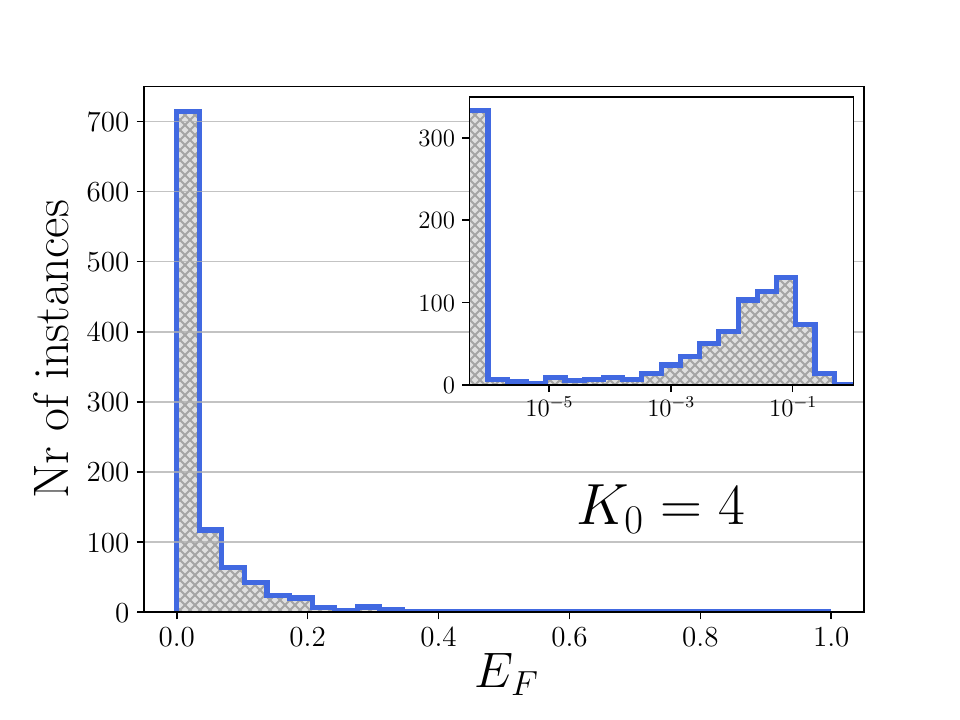} 
	\caption{Number of states with entanglement of formation $E_F$, given a set of $1000$ random instances, for $N=2$ and $\nstates_0 = \nstates = 2,3,4$. 
	Insets display the x-axes in logarithmic scale, to highlight the distribution states with low entanglement content.
	Notice that almost separable states are more typical for larger $\nstates_0$. 
	 }
	 \label{SM:fig:RandPS_N2}
\end{figure}

Let us start with $N=2$ qubits, a case for which the exact \eof can be computed. 
We study ensembles of $\nstates_0 = 2, 3$ and $4$ pure states, fixing $\nstates = \nstates_0$. The corresponding results are outlined in Fig.~\ref{SM:fig:RandPS_N2}. 
For $\nstates_0=2$ the density matrix can still have a considerable amount of entanglement and no separable states are found. 
On the other hand, for $\nstates_0=4$ we observe that the typical state is almost separable. This is highlighted in the insets of Fig.~\ref{SM:fig:RandPS_N2}, where the x-axis is given in logarithmic scale. 
We benchmarked our optimization instance by instance against the corresponding exact results from concurrence. Among 1000 samples with $\nstates_0=4$, all cases present an absolute error $< 10^{-3}$, which is even $< 10^{-6}$ for $99\%$ of them. 

\begin{figure}
	\centering
	\includegraphics[width=\columnwidth]{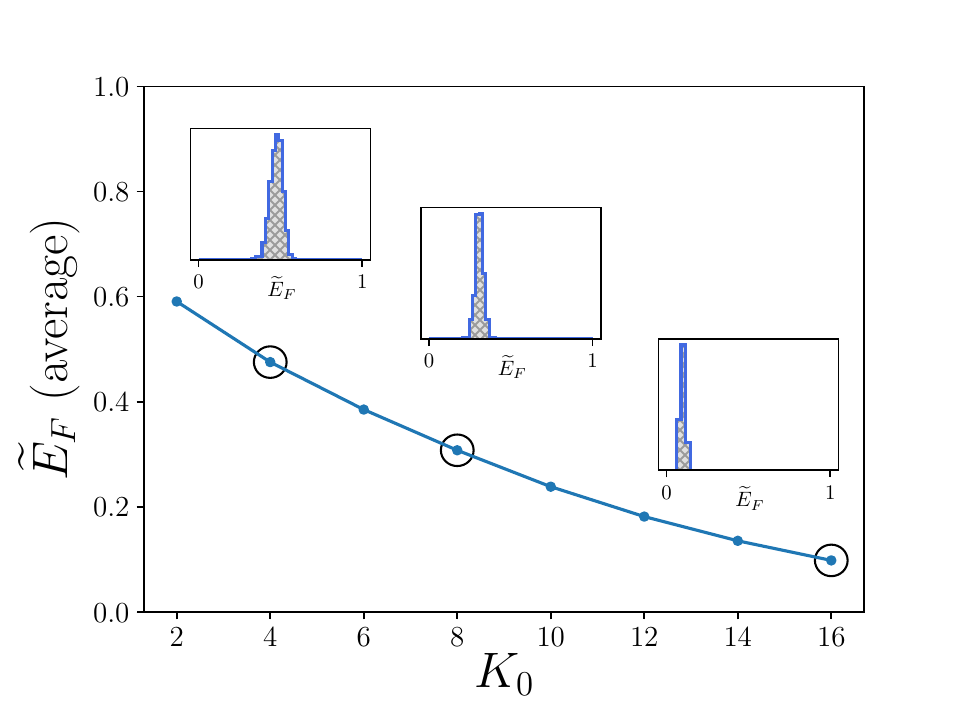}
	\caption{$\widetilde{E}_F = E_F / 2$ \textit{versus} $\nstates_0 = \nstates$, for $N=4$ qubits. Points are the mean of the distributions for random ensembles at fixed $\nstates_0$. Notice that all errors on the mean are smaller than the marker size. 
	Insets: distribution of the entanglement for $\nstates_0 = 4,8,16$. 
	}
	\label{SM:fig:RandPS_vschi0_N4}
\end{figure}

Let us now consider states of $N>2$ qubits, for which no exact solution exists. 
We first fix the number of qubits to $N=4$ and study how entanglement changes as a function of $\nstates_0$, the number of random pure states in the ensemble. 
To focus on a size-independent quantity, we study the entanglement density $\widetilde{E}_F = E_F / (N/2)$, with $N/2$ being the entanglement of a maximally entangled state of $N$ qubits with bipartitions of the same dimension. 
Fig.~\ref{SM:fig:RandPS_vschi0_N4} shows that adding more and more random states to the ensemble typically leads to a smaller entanglement density. 
For each $\nstates_0 = \nstates$, we plot the mean of the distribution of the entanglement computed from several random ensembles. For every point, we find the error on the mean is smaller than the marker size, so it is not visible. Indeed, all the distributions are quite peaked, as shown in the insets for the $\nstates_0 = 4,8,16$ cases. 

\begin{figure}
	\centering
	\includegraphics[width=\columnwidth]{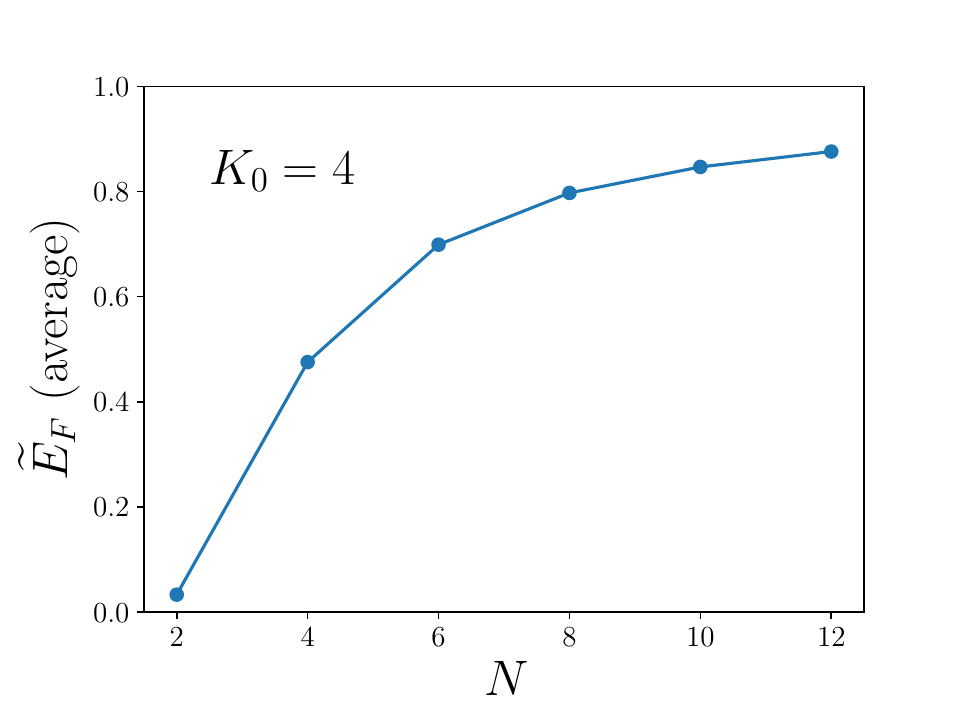}
	\caption{Average entanglement density \textit{versus} the number of qubits $N$, for $\nstates_0 = \nstates = 4$. 
	Notice that here $\widetilde{E}_F$ increases with $N$, because $\nstates_0 / \dimH$ decreases for increasing number of qubits. 
	}
	\label{SM:fig:RandPS_vsN_chi4_time}
\end{figure}

It is also interesting to fix $\nstates_0=4$ and vary $N$. In this case, 
the entanglement density 
grows as the number of qubits increases, as shown in Fig.~\ref{SM:fig:RandPS_vsN_chi4_time} (again we set $\nstates_0=\nstates$). 
This, together with the previous results in Fig.~\ref{SM:fig:RandPS_vschi0_N4}, suggests that entanglement between qubits is typically stronger for smaller values of the ratio between $\nstates_0$ and the Hilbert space dimension, $\nstates_0 / \dimH$. 

\subsubsection{Structured random ensembles}
\label{sec:structured_random}
We consider here random density matrices distributed according to the Hilbert-Schmidt measure~\cite{Zyczkowski_JPA01,Zyczkowski_JPA04,GeneratingRandomDM_JMP11}. 
They can be generated from the following procedure~\cite{GeneratingRandomDM_JMP11}: 
\begin{enumerate}
	\item take a random complex square matrix $X$ belonging to the so-called Ginibre ensemble, \ie  with real and imaginary parts of each component chosen independently according to the normal distribution; 
	\item construct the state 
	\begin{equation}	\label{SM:eq:XXdag}
		\rho = \frac{ X X^\dagger }{ \trace{X X^\dagger} } \; . 
	\end{equation}
\end{enumerate}
%
We then obtain the initial representation $X$ by exact diagonalization of $\rho$. 
This class of density matrices are typically full rank, so that $\nstates_0 = \dimH$. In the following, we also fix $\nstates = \nstates_0$. 

\begin{figure}
	\centering
	\includegraphics[width=\columnwidth]{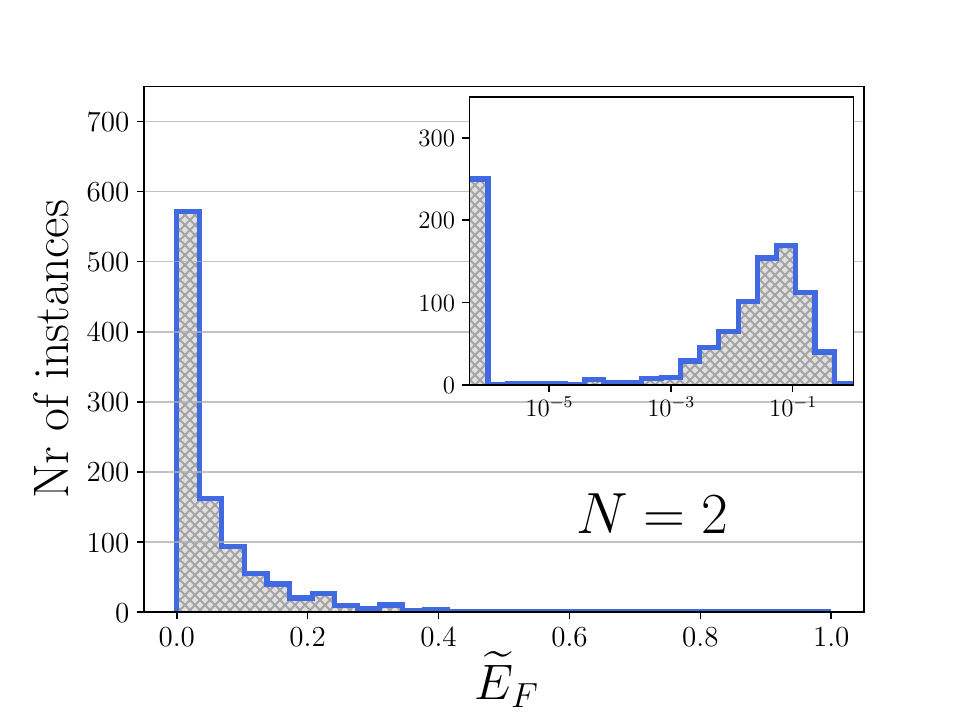} \\
	\includegraphics[width=\columnwidth]{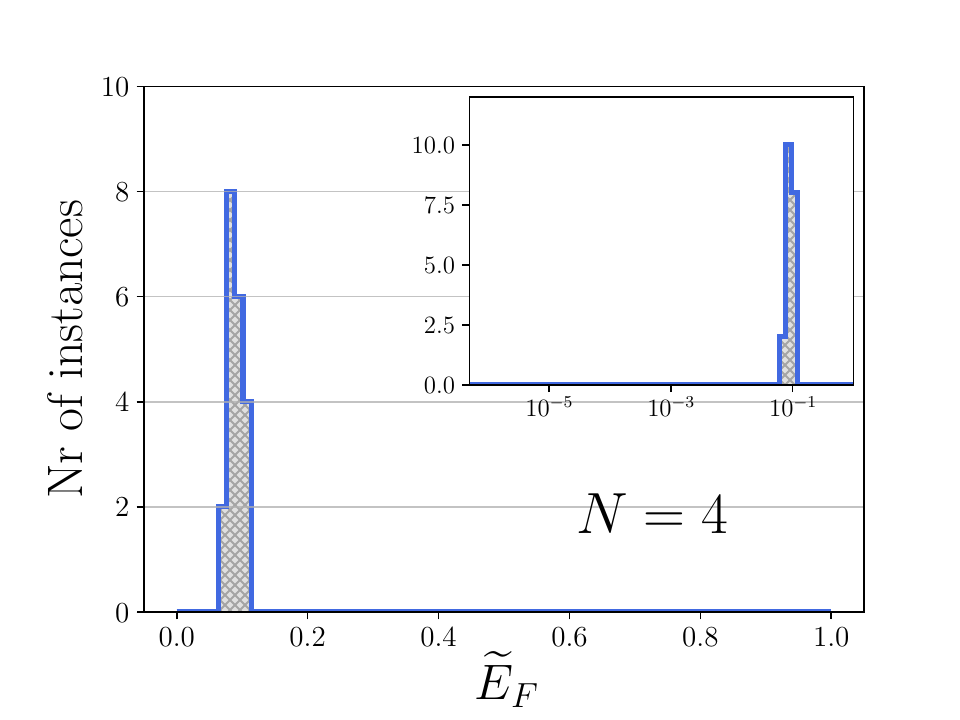}
	\caption{ Distribution of the entanglement density over random density matrices for $N=2$ (top) and $N=4$ (bottom) qubits, sampled as outlined in Sec.~\ref{sec:structured_random}. 
	Top: 
	distribution from 5000 two-qubits states. For each instance, we compared the entanglement with the exact result from concurrence: all absolute errors are below $10^{-6}$. 
	Bottom: distribution from 20 instances, very peaked at 
	$\widetilde{E}_F = E_F/2 \sim 0.1$. 
	}
	\label{Res:fig:RandDM_histo}
\end{figure}

Let us start with two qubits, $N=2$. We computed the \eof for 5000 random samples belonging to this class, both exactly through concurrence and by our optimization approach. 
Results from each instance are all in perfect agreement, since absolute errors are below $10^{-6}$ (not shown). 
The distribution of the entanglement for such states is given in Fig.~\ref{Res:fig:RandDM_histo} (top). 
We observe that these states are typically little (or not) entangled and that the number of separable states is around $25\%$ of the total. 

We also investigated ensembles of $N=4$ qubits. Since $\nstates_0=16$ in this case, the optimization task is more complex and we therefore have a much smaller set of instances. 
Nevertheless, the results clearly show a very peaked distribution at around $\widetilde{E}_F = E_F / 2 \sim 0.1$, see Fig.~\ref{Res:fig:RandDM_histo} (bottom).

\subsubsection{Random separable states}
In this section, we test the reliability of the minimization in finding zero entanglement on random separable mixed states. 
Density matrices are assumed to have the form
\begin{equation}
	\rho = \sum_{i=1}^{\nstates_0} \; p_i \; \rho_i^A \otimes \rho_i^B \; ,
\end{equation}
where $\nstates_0$ is a random integer $\in [1,4]$ extracted according to a uniform distribution, 
probabilities $p_i \in [0,1]$ are also chosen according to a uniform distribution, and the random density matrices $\rho_i^A$, $\rho_i^B$ are independently constructed following the same procedure used for structured random ensembles, see Eq.~\eqref{SM:eq:XXdag}. 

\begin{figure}
	\centering
	\includegraphics[width=\columnwidth]{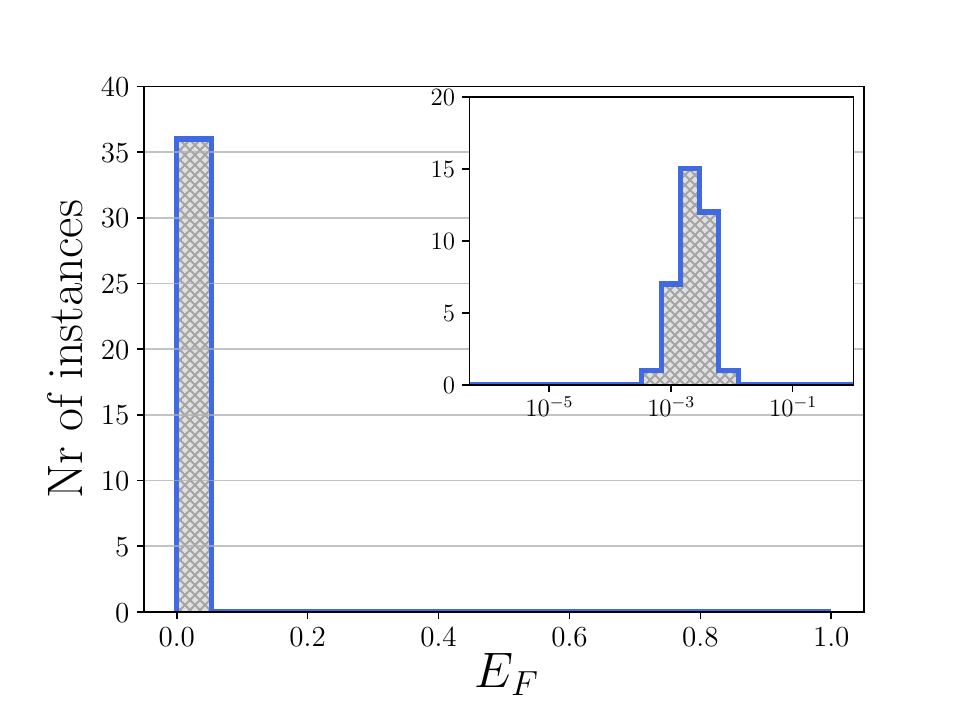}
	\caption{Distribution of the \eof for 36 instances of random separable states of $N=4$ qubits. These states are typically full rank, so that $\nstates = \nstates_0 = 16$. 
	Separability is satisfactorily assessed, $E_F \lesssim 10^{-2}$. 	
	}
	\label{SM:fig:separable}
\end{figure}

Given 10000 samples for $N=2$ qubits, \eof has always been found to be $< 10^{-6}$ (not shown). 
The same test for $N=4$ qubits is instead shown in Fig.~\ref{SM:fig:separable}, where we see that all 36 instances have very low entanglement $\lesssim 10^{-2}$. However, we cannot determine separability as clearly as for the $N=2$ case, because in this case the optimization problem is much harder ($\nstates_0 = 16$).

\subsubsection{Werner states}
\begin{figure}
	\centering
	\includegraphics[width=\columnwidth]{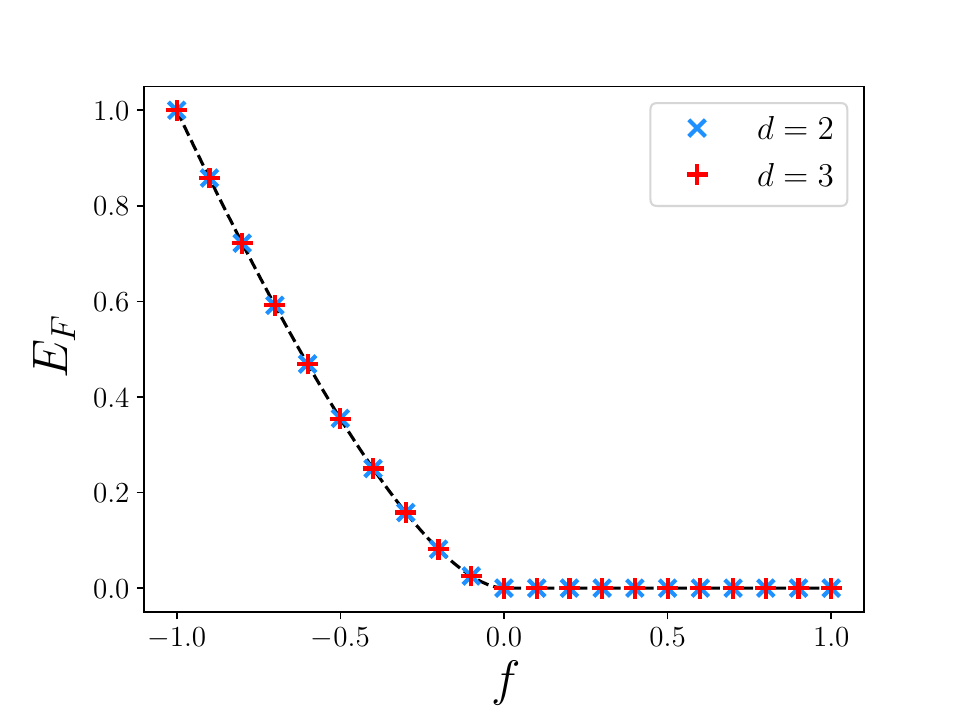} 
	\caption{Werner states: $E_F$ \textit{versus} $f = \trace{\mathbb{F}\rho}$ for two qubits ($d=2$) and two qutrits ($d=3$). The black dashed line is the exact result, which is independent of $d$. 
	In some instances, especially for $f \geq 0$, local minima for the \eof are present. We nevertheless can find the correct results after repeating the calculation at most 3 times with $K = K_0$. 
	The point at $f=1$ for $d=3$, corresponding to $K_0=6$, required instead $K=9$ to get the correct result. 	
	 }
	\label{SM:fig:Werner}
\end{figure}
Werner states form the class of bipartite quantum states invariant under $U \otimes U$ transformations, where $U$ is a unitary matrix acting on one bipartition, with the two bipartitions assumed to have the same Hilbert space dimension $d$~\cite{Werner_PRA89, Werner_PRA01}. 
They are defined as 
\begin{equation}
	\rho = \frac{1}{d(d^2 - 1)}	\Big[ (d-f) \id + (df-1)\mathbb{F} \Big] \; , 
\end{equation}
where $\id$ is the identity operator on one bipartition, $\mathbb{F} = \sum_{ij} \ketbra{ij}{ji}$ is the so-called flip operator 
and $f = \trace{ \mathbb{F} \rho }$. 
Whenever $f>0$ the state is separable, while for $f \leq 0$ the exact solution is known and independent of $d$~\cite{Werner_PRA89, Werner_PRA01}. 
%

%
%

From Fig.~\ref{SM:fig:Werner} we see that the optimizations for two qubits and two qudits have perfectly converged towards the actual values (black dashed line) using the Nelder-Mead method. 
For the two-qutrits calculations, we observe presence of local minima for the \eoff, where the minimizer remains trapped. However, repeating the same calculation few times, starting from different initial parameters, returns the correct result. 
We have set $K = K_0$ for all points, with the exception of $d=3, f=1$, where we needed $K = K_0+3=9$.

\subsubsection{Isotropic states}
Analogously to Werner states, isotropic states are obtained as the class of states invariant under $U \otimes U^*$ transformations, where $U^*$ is the complex conjugate of $U$~\cite{Isotropic_PRA99}. They are defined as 
\begin{equation}
	\rho = \frac{1-f}{d^2-1} \big( \id - P_+ \big) + f P_+ \; ,
\end{equation}
where $d$ is the Hilbert space dimension of one of the two identical bipartitions, 
$P_+ = \ketbra{\psi_+}{\psi_+}$ is the projector onto the maximally entangled state 
$\ket{\psi_+} = \frac{1}{\sqrt{d}} \sum_{i=1}^d \ket{ii}$,
\begin{figure}[H]
	\centering
	\includegraphics[width=\columnwidth]{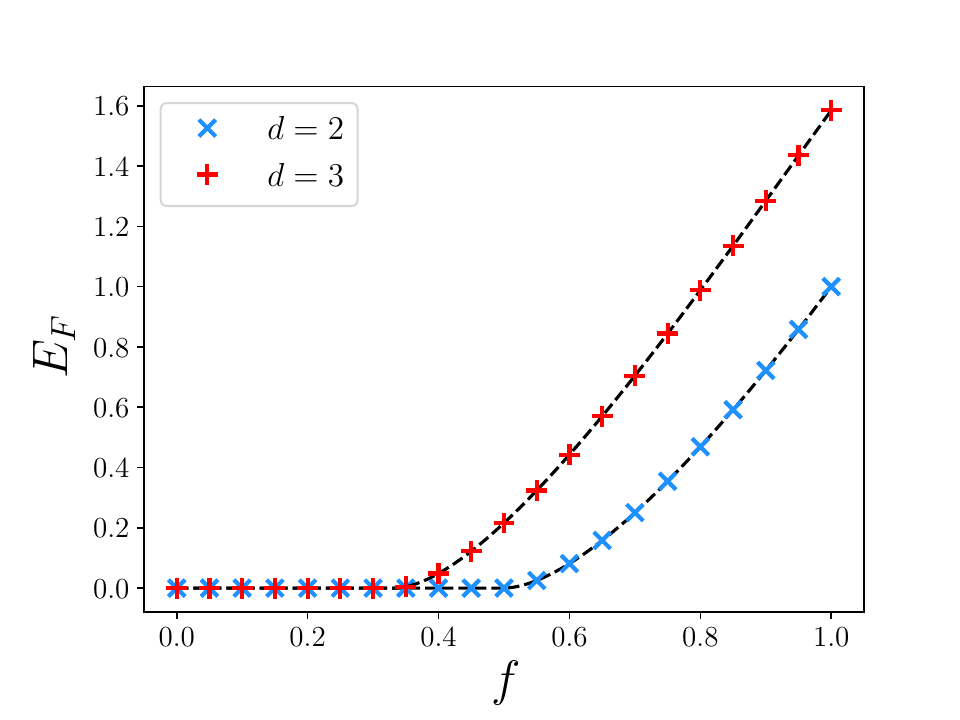} 
	\caption{Isotropic states: $E_F$ \textit{versus} the fidelity $f$ with the maximally entangled state, for two qubits ($d=2$) and two qutrits ($d=3$). 
	The two black dashed lines correspond to the exact results for $d=2,3$. 
	Perfect agreement is found as for Werner states, where few minimization repetitions were needed for separable states to achieve the correct results. 
	We set $K = K_0$ everywhere, with the exception of $d=3, f=0$, which required $K = K_0 + 1 = 9$. 
	}
	\label{SM:fig:Isotropic}
\end{figure}
\noindent
and 
$f = \expctval{\psi_+}{\rho}{\psi_+}$. 
Also in this case, the \eof can be computed exactly for any $d$~\cite{Terhal_PRL00}: 
isotropic states are separable for $0 \leq f \leq 1/d$, while they are entangled for $1/d < f \leq 1$. 

In Fig.~\ref{SM:fig:Isotropic}, we compare the results from our approach against the exact solution: as already found for Werner states, the Nelder-Mead algorithm works very well for both the two-qubits ($d=2$) and the two-qutrits ($d=3$) cases. 
The latter required however few repetitions of the same minimizations, especially when dealing with separable states. 

\end{document}